\documentclass[aps,prl,twocolumn,floatfix,showpacs]{revtex4}
\usepackage{amsmath,amssymb,graphics}
\begin{document}
\title{Quantum Monte Carlo calculations of structural properties of
  FeO under pressure}

\author{Jind\v rich Koloren\v c}
\altaffiliation[On leave from ]{Institute of Physics, Academy of
  Sciences of the Czech Republic, Na Slovance~2, CZ-18221 Praha~8,
  Czech Republic}
\email[]{kolorenc@fzu.cz}
\author{Lubos Mitas}
\affiliation{Department of Physics and CHiPS, North Carolina State University,
  Raleigh, North Carolina 27695, USA}

\date{\today}

\begin{abstract}
We determine the equation of state of stoichiometric FeO employing the
diffusion Monte Carlo method. The fermionic nodes are fixed to those of a
wave function having the form of a single Slater determinant. The
calculated ambient pressure properties (lattice constant, bulk
modulus and cohesive energy) agree very well with available
experimental data. At approximately 65 GPa, the lattice structure
is found to change from rocksalt type (B1) to NiAs based (inverse B8).
\end{abstract}

\pacs{ 72.80.Ga, 71.15.-m, 71.20.-b, 64.30.+t }


\maketitle

Transition metal oxides are solids with strong
electron-electron correlations that lead to a rich variety of observed
structural and electronic phases. Almost all transition metal oxides
are a problem on its own due to competitive interplay among correlation
and exchange in $d-$subshells, crystal fields, $d-$$p$ hybridization and
charge transfer. Elucidation of high-pressure properties of FeO is of
particular interest in geophysics, since this compound belongs to
constituents of the deep Earth's interior. FeO is also one of the
challenging simple oxides due to the nominally open-shell occupation
of the $3d$ levels.  Indeed, this system proved to be problematic for
the density functional theory (DFT) in its local density or
generalized gradient approximations (LDA or GGA). For example, both
LDA and GGAs predict an incorrect ground state lattice structure
\cite{mazin1998,fang1998,fang1999}.


At ambient conditions, FeO crystallizes in B1 (NaCl-type)
structure. It is antiferromagnetically ordered at temperatures below
198 K and this ordering is accompanied by a small rhombohedral
distortion denoted as rB1---the unit cell is stretched along the [111]
body diagonal.  In shock-wave studies it was observed that around
70~GPa the oxide transforms to a different structure
\cite{jeanloz1980}, which was inferred as B2 (CsCl-type) in analogy
with similar materials, but LDA calculations hinted that much larger
pressure, around 500 GPa, would be needed to stabilize B2 against the
B1 phase \cite{isaak1993}. Besides that, no such structural transition
was detected in static compression experiments \cite{yagi1985}, unless
the material was significantly heated up \cite{fei1994}. X-ray
diffraction performed along the high-temperature static compression
revealed that the high-pressure structure is actually B8 (NiAs-type)
\cite{fei1994}.

There are two distinct ways of putting FeO on NiAs lattice, the so-called
normal B8, where iron occupies Ni sites, and inverse B8 (iB8 for short),
where iron sits on As sites. It is the latter configuration that comes
from band-structure theories as the more stable of the two
\cite{mazin1998,fang1998,fang1999}. Also, reinterpretation of the data of
Ref.~\onlinecite{fei1994} suggests that the high pressure phase
is a mixture of B8 and iB8 phases \cite{mazin1998}. On the theoretical
side, the introduction of the iB8 structure into the picture revealed
a serious deficiency in the LDA (and GGA) as applied to FeO, since
the iB8 phase is predicted more stable not only than B8 but also than
B1 at all pressures, which contradicts experimental findings. It
was demonstrated that inclusion of Coulomb $U$ to better account
for electron-electron correlations alleviates this problem
\cite{fang1998,fang1999}.


In this Letter, we calculate the equation of state of stoichiometric FeO
using the fixed-node diffusion Monte Carlo method (DMC)
\cite{foulkes2001}, a many-body computational technique
that accurately treats even strongly correlated
systems. Based on the aforementioned studies, we confined ourselves 
to only two structures---B1 with the type-II antiferromagnetic (AFM)
ordering (symmetry group R$\bar{3}$m) and iB8 also in the AFM state
(group P$\bar{6}$m2).
We show that DMC, even in its simplest version
based on a single-determinant 
Slater--Jastrow wave function, provides a
very consistent picture of this complicated system that closely
follows experimental data including estimation of the transition
pressure. 

The guiding wave function that defines (fixes) the fermionic nodes 
in our DMC simulations is of the Slater--Jastrow type
$\Psi_G=\Psi_S \exp[J]$, where
\begin{subequations}
\label{eq:wf}
\begin{align}
\label{eq:wf_sj}
\Psi_S(\mathbf{r}_1,\dots,\mathbf{r}_N) &=\det\{\psi_{\sigma}\}=
\det\{\phi_{\alpha}^{\uparrow}\}
\det\{\phi_{\beta}^{\downarrow}\}\,,\\
\label{eq:wf_s}
J(\mathbf{r}_1,\dots,\mathbf{r}_N) &=
    \sum_{i,j} f(\mathbf{r}_i-\mathbf{r}_j)
  + \sum_{i,I} g(\mathbf{r}_i-\mathbf{R}_{I})\,.
\end{align}
\end{subequations}
The lower-case indices in Eq.~\eqref{eq:wf_s} run over electrons,
while the upper-case index denotes ions.  The Jastrow correlation
factor~$J$ contains one- and two-body terms, $g$ and $f$, that have
the same form as in Ref.~\onlinecite{wagner2007} and represent 17
variational parameters that were optimized within variational Monte
Carlo (VMC) framework. The single determinant of spinorbitals
$\psi_{\sigma}$ becomes a product of spin-up and -down determinants of
spatial orbitals $\{\phi_{\alpha}^{\uparrow},
\phi_{\beta}^{\downarrow}\}$ after fixing the electron spins,
$N^{\uparrow}=N^{\downarrow}=N/2$, while the overall state is a
spin-unrestricted antiferromagnet.

The large energy scale of core electrons poses a difficulty to the DMC
in an analogous way as it does to plane-wave based electronic
structure techniques. Therefore, we replace the atomic cores by
norm-conserving pseudopotentials \cite{lee_private_compact} within the
so-called localization approximation \cite{mitas1991}. We utilize only
small-core pseudopotentials to minimize losses in accuracy as much as
possible. We have argued recently \cite{kolorenc2007} that even
small-core pseudopotentials could lead to imprecisions in description
of transition metal compounds if a spin-related transition occurs as a
part of the phenomena of interest. However, it should not affect the
present calculations, since iron atoms in FeO stay in a high-spin
state in the whole range of pressures we study in both B1 and iB8
phases. This applies to the DMC results as well as to the flavors of
DFT that we used to construct the Slater determinants.

The quality of the fixed-node DMC total energy is determined solely by
the quality of fermionic nodes of the guiding wave function. When the
form of Eq.~\eqref{eq:wf} is adopted, the parameters controlling
location of fermionic nodes are the one-electron
orbitals~$\{\phi_{\alpha},\phi_{\beta}\}$. The direct VMC optimization
of these orbitals
is currently impractical for simulation sizes required in the present
study.  Instead, we use one-electron orbitals from spin-unrestricted
calculations with hybrid PBE0 functional given as \cite{perdew1996b}
\begin{equation}
\label{eq:pbe0}
E^{PBE0}_{xc}=a E_x^{HF} + (1-a) E^{PBE}_{x} + E^{PBE}_c\,.
\end{equation}
Here $E^{PBE}_{x}$ and $E^{PBE}_c$ are exchange and correlation parts
of the PBE-GGA \cite{perdew1996}, $E_x^{HF}$ is the exact exchange
from Hartree--Fock (HF) theory and the weight $a$ is in the range
$0<a<1$. We have found that the inclusion of exact exchange term into
PBE-GGA has similar effect as Coulomb~$U$ in the LDA$+U$ method. It
opens a gap in the electronic spectrum of the AFM B1 phase and
stabilizes it relative to the iB8 structure. Both the gap and the
transition pressure increase with increasing~$a$. With mixing weight
$a=0.05$, the iB8 is still more stable than B1 everywhere, the B1 to
iB8 transition occurs at 5 GPa for $a=0.1$ and at 43 GPa for
$a=0.2$. Note that in MnO, which exhibits similar structural transition,
the experimental range of transition pressures is reached already for
$a\approx 0.1$ \cite{kolorenc2007}, while in the present case of FeO,
the transition takes place much sooner than in experiments ($\geq 70$
GPa) even for twice as much exact-exchange content in PBE0.

The exchange-correlation functional of Eq.~\eqref{eq:pbe0} defines
one-parametric class of single-particle
orbitals~$\{\phi_{\alpha}^{(a)}, \phi_{\beta}^{(a)}\}$, which can be
used to minimize the DMC fixed-node error by varying the exact
exchange weight~$a$.  Although in simple insulators, such as silicon,
the differences between fixed-node energies corresponding to various
sets of one-particle states were found to be rather marginal
\cite{kent1998}, more pronounced 
differences have been obtained for transition metal compounds.
In isolated molecules of transition metal monoxides, the fixed-node
DMC energies with orbitals from B3LYP (hybrid functional similar to
PBE0) are noticeably 
lower than with HF or pure DFT orbitals
\cite{wagner2003,wagner2007}. DMC optimization of the exact exchange
proportion in the B3LYP was performed in Ref.~\onlinecite{wagner2003}
for the MnO molecule. The optimal value was reported as approximately
17\%, but the minimum was rather broad and shallow and values between
5\% and 30\% were almost equivalent. Therefore, we have chosen the
weight~$a$ in PBE0 to be $a=0.2$, corresponding to 20\% of GGA
exchange being replaced with the exact exchange. In the following we
abbreviate this functional as PBE0${}_{20}$. This choice is compatible
with findings of Ref.~\onlinecite{wagner2003} and leads to reasonable
ambient-pressure properties of FeO already within (hybrid) DFT, i.e.,
B1 structure is insulating and more stable than iB8. We also checked
that at equilibrium the PBE0${}_{20}$ orbitals provide DMC energy 0.3
eV per FeO lower than orbitals from the HF approximation.

In our simulations, the infinite crystal was modelled by a
periodically repeated simulation cell containing 8 FeO units, i.e.,
176 valence and semi-core electrons. Although such a system is
certainly not small to deal with in an explicitly many-body fashion,
it turns out that finite-size effects are significant if not treated
properly. The origin of these finite-size errors is twofold. One part
is related to incorrect momentum quantization due to confinement of
electrons into the simulation cell, the second part comes from the
artificial periodicity of exchange-correlation hole due to periodic
extension of Coulomb potential using the Ewald summation.

The problem associated with confinement appears also in mean-field
band theories, where it is exactly resolved by integration over the
first Brillouin zone, while working only within the smallest
``simulation'' cell possible, the primitive cell. Each ${\bf k}$-point
in the first Brillouin zone corresponds to a different boundary
condition
imposed on the primitive cell. The momentum integration is equivalent
to averaging over these so-called twisted boundary
conditions. Analogous averaging procedure performed in a many-body
simulation does not represent a complete correction, but it proved to
be very efficient within Monte Carlo algorithms \cite{lin2001}.
In this study we deal only with insulating states, which 
simplifies matters considerably and  we have found that just 8 twists are
enough for our simulation cell size. We have verified that within 
PBE0${}_{20}$ the total energy obtained in our simulation cell using
just 8 ${\bf k}$-points differs only $\approx 0.01$ eV per
FeO from fully converged integral over Brillouin zone.

\begin{figure}
\resizebox{\linewidth}{!}{\includegraphics{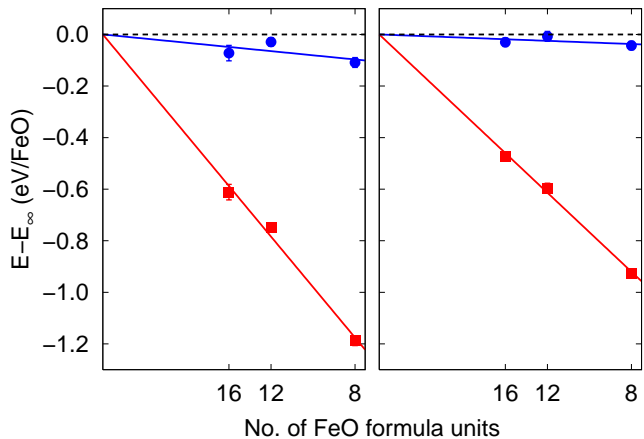}}
\caption{\label{fig:fss}(color online) Finite-size
  errors of the twist averaged DMC energy
  at volumes $15.9$ \AA${}^3\!/$FeO (left) and
  $20.4$ \AA${}^3\!/$FeO (right). Pure Ewald energies are shown as
  red squares, values corrected according to Eq.~\eqref{eq:sk_corr}
  are represented by blue circles. Note that the finite-size errors
  increase with increasing the electron density. The energy in
  infinite cell $E_{\infty}$ is extrapolated from the data shown.
  Statistical errorbars are smaller than symbol sizes except for the
  largest cell in the left panel.}
\end{figure}

The second part of finite-size errors, originating from artificial
periodicity of the Ewald potential, is accounted for with the aid of the
correction introduced in Ref.~\onlinecite{chiesa2006}. The estimate
for the energy at infinite volume is written as 
\begin{equation}\label{eq:sk_corr}
E=E_{Ewald}+\frac1{4\pi^2}\int_D
    d^3\!k\,\frac{S(\mathbf{k})}{k^2}\,.
\end{equation}
The static structure factor is defined as
$S(\mathbf{k})=\langle\Psi_0|\hat\rho_{\mathbf{k}}
\hat\rho_{-\mathbf{k}}|\Psi_0\rangle/N$ with $\hat\rho_{\mathbf{k}}$
standing for a Fourier component of the electron density. The integral in
Eq.~\eqref{eq:sk_corr} runs over a domain $D$
centered around $\mathbf{k}=0$ and having volume $8\pi^3/\Omega$, where
$\Omega$ is volume of the simulation cell. The structure factor
$S(\mathbf{k})$ is evaluated within the DMC at a discrete set of points
and then extrapolated towards $\mathbf{k}=0$. Performance
of the correction given by Eq.~\eqref{eq:sk_corr} applied to FeO
is illustrated in Fig.~\ref{fig:fss}, where we plot the total energy
at two different electron densities as calculated in simulation cells
of varied size up to 16 FeO units, i.e., 352 valence and semi-core
electrons. The correction removes more than 90\% of the finite-size
error introduced by the periodic electron-electron interaction
potential and enables us to replace
expensive size scaling analysis of Fig.~\ref{fig:fss} by a
simple formula, Eq.~\eqref{eq:sk_corr}.

{\itshape Results.} First we discuss properties of the B1 phase around
equilibrium volume. 
For the cohesion energy, $E_{coh}=E_{\rm Fe}+E_{\rm O}-E_{\rm FeO}$,
our DMC simulations yield $9.66\pm 0.04$ eV/FeO that matches $9.7$
eV/FeO deduced from experimental formation enthalpies \cite{crc}. The
electronic gap, which we calculate as a difference between total
energies of the ground state and the first excited state at the
$\Gamma-$point in our 8 FeO simulation cell, comes out as $2.8\pm 0.3$
eV.
This value is not too far from optical absorption edge observed near
$2.4$ eV \cite{bowen1975}. The weak feature displayed between
$1.0$ and $1.5$ eV in these experiments is not reproduced in the
picture of FeO we present here. However, it is quite possibly related
to imperfections in structural or magnetic order, since essentially
the same absorption band was repeatedly observed in other systems
where Fe atoms act as impurities \cite{jones1967,hjortsberg1988}.

\begin{figure}
\resizebox{\linewidth}{!}{\includegraphics{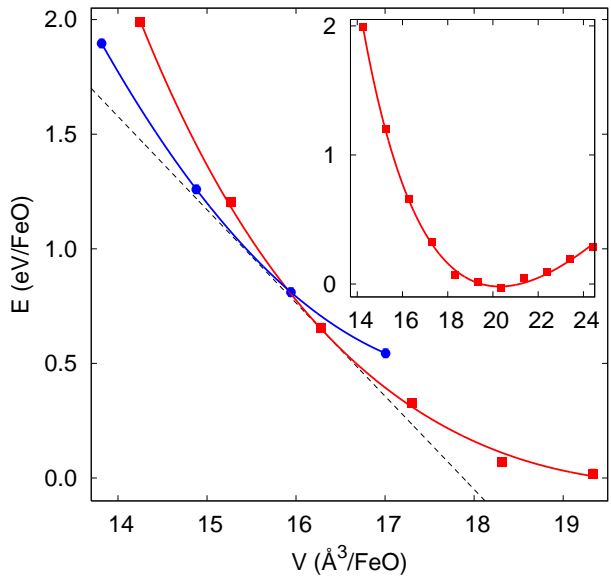}}
\caption{\label{fig:ev}(color online) Total energies of
  B1 (red squares) and iB8 (blue circles) phases. Statistical
  errorbars are smaller than symbol sizes. Lines are
  least-square fits with Murnaghan equation of state. The dashed
  double tangent corresponds to the transition pressure of
  65 GPa. Inset shows the data for B1 phase over wider volume
  region, including equilibrium.}
\end{figure}
\begin{table}[b]
\caption{\label{tab:ambient}Equilibrium lattice constant $a_0$, bulk
  modulus $K_0$ and its derivative $K_0'=(\partial K_0/\partial P)_T$
  calculated in this work (DMC and PBE0${}_{20}$) compared to selected
  theories and experiments. The experimental lattice constant is
  extrapolated to the stoichiometric FeO, whereas the values of $K_0$
  and $K_0'$ are not. The extrapolated value of bulk modulus $K_0$ is
  estimated around 180 GPa \cite{mccammon1984,jackson1990}.}
\begin{tabular}{l@{\extracolsep{2em}}ccc}
\hline\hline
                        & $a_0$ (\AA) & \ $K_0$ (GPa)\  & $K_0'$ \\
\hline
DMC                     & 4.324(6)    & 170(10)     &  5.3(7) \\
PBE0${}_{20}$           & 4.328       & 182         & 3.7    \\
GGA \cite{fang1999}     & 4.28        & 180         & 3.6 \\
LDA \cite{isaak1993}    & 4.136       & 173         & 4.2 \\
experiment\ \           & 4.334 \cite{mccammon1984}
                        & 152.3 \cite{jackson1990}
                        & 4.92 \cite{jackson1990}\\
\hline\hline
\end{tabular}
\end{table}

The DMC energy as a function of volume, together with fitted Murnaghan
equation of state, is presented in Fig.~\ref{fig:ev}.  The parameters
of the least-square fit are compared with predictions of other
electronic structure methods and with experiments in
Tab.~\ref{tab:ambient}. The DMC estimate for equilibrium lattice
constant~$a_0$ is in excellent agreement with experimental value
extrapolated to stoichiometric FeO and offers more consistent
prediction than LDA or GGA. On the other hand, the hybrid
PBE0${}_{20}$ functional, which we used to construct the DMC guiding
wave function, provides a similar value.  All methods shown in
Tab.~\ref{tab:ambient} provide essentially the same value of bulk
modulus, which is noticeably larger than typical experimental
reports. The extrapolation to stoichiometry is, however, expected to
lead to values in the vicinity of the theoretical data
\cite{mccammon1984,jackson1990}. The isothermal pressure derivative of
the bulk modulus, $K_0'$, turns out to be larger in DMC than in the
density-functional approaches, which makes it compatible with
elastic-wave experiments \cite{jackson1990}.

The equation of state calculated within diffusion Monte Carlo up to
large hydrostatic pressures is shown in 
Fig.~\ref{fig:ev}. The $c/a$ ratio in the hexagonal iB8 phase, stable
at high pressures, was optimized within PBE0${}_{20}$. It was found
to increase from 1.93 at volume 17 \AA${}^3\!$/FeO to 2.03 at
14 \AA${}^3$\!/FeO. These ratios agree well with experimental
$c/a=2.01$ at $14.83$ \AA${}^3\!/$FeO reported in
Ref.~\onlinecite{fei1994}.

The B1 phase, stable at low pressures, was assumed cubic, i.e., we
neglected the rhombohedral distortion. 
We did not use the DFT optimized geometries in this case, because DFT
based techniques are not conclusive in determination of the
rhombohedral distortion \cite{cococcioni2005,gramsch2003,zhang2007}.
We checked the impact of fixing the cubic symmetry by comparing DMC
total energies for different distortions at high compression, where the
distortion has the largest impact.  At the volume of $15.3$
\AA${}^3\!$/FeO, the energy gain associated with the rhombohedral
distortion was of the order of statistical errorbars $\approx 0.02$
eV$\!$/FeO, i.e., too small to affect the results.

The critical pressure $P_c$ of the structural transition from B1 to
iB8 phase was determined from equality of Gibbs potentials,
$G_{B1}(P_c)=G_{iB8}(P_c)$, graphical equivalent of
which---the double tangent---is shown in Fig.~\ref{fig:ev}. The
value is $P_c=65\pm 5$ GPa with the errorbar given by statistical
fluctuations of the Monte Carlo simulations. 
The prediction $P_c=65$ GPa agrees quite well with
shock-wave data and with high-temperature static compression
experiments, except for the fact that our investigation is 
performed at $T=0$, for which experiments suggest considerably higher
transition pressure. Our finding could be interpreted as an indirect
support for existence of an energy barrier between the two phases that
requires a thermal activation to be overcome.

\begin{figure}
\resizebox{\linewidth}{!}{\includegraphics{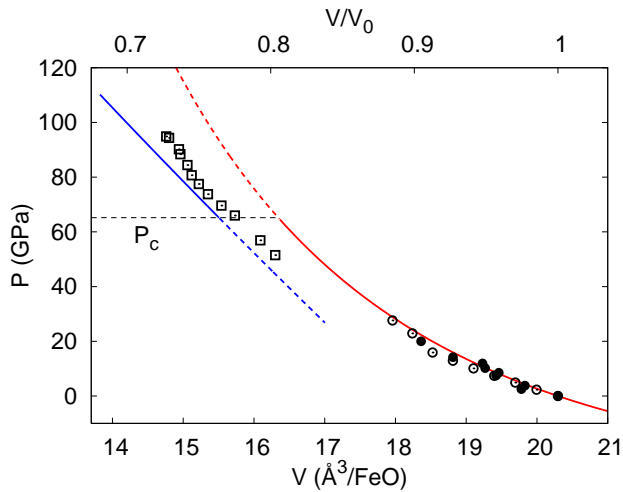}}
\caption{\label{fig:pv}(color online) $P(V)$ curves of B1 (red) and iB8
  (blue) phases. Lines are our DMC fits as in Fig.~\ref{fig:ev}. Points
  are   experimental data: filled circles 
  \cite{yagi1985} (B1, Fe${}_{0.98}$O), empty  
  circles \cite{clendenen1966} (B1, Fe${}_{0.94}$O), empty squares
  \cite{fei1994} (iB8, Fe${}_{0.98}$O). All B1 data
  \cite{yagi1985,clendenen1966} were taken at room temperature, the
  iB8 data \cite{fei1994} correspond to 900 K. B1 datasets are
  shown relative to the equilibrium volumes reported in the individual
  studies to approximately remove the non-stoichiometry effects.}
\end{figure}

Another means of comparison with experiments is looking at the $P(V)$
equation of state, Fig.~\ref{fig:pv}. Agreement
between data for B1 structure (after extrapolation to stoichiometric
FeO) is very good, which is in concord with similarly good
correspondence of ambient pressure parameters compared in
Tab.~\ref{tab:ambient}. Our curve for iB8 structure also follows the
x-ray data of Ref.~\onlinecite{fei1994} rather nicely (no
stoichiometry related correction was attempted in this case). Experimentally,
this phase is perhaps somewhat stiffer than in our calculations, which
signals that DMC is likely to slightly underestimate the transition
pressure~$P_c$.

{\itshape In summary}, the equation of state and basic
electronic structure of FeO calculated with
the diffusion Monte Carlo agrees very well with many aspects of
available experimental data. Considering that the DMC is essentially a
parameter-free method and that the fixed-node condition was enforced
with the aid of a very simple wave function (single Slater
determinant), the degree of consistency of the provided picture is
quite remarkable. 

We acknowledge support by NSF DMR-0121361 and EAR-0530110
grants. This study was enabled by INCITE allocation at ORNL and by
allocation at NCSA.  QMC simulations were done using \textsc{QWalk}
code \cite{qwalk}, and the one-particle orbitals 
were calculated with \textsc{Crystal2003}
\cite{crystal2003}.





\end{document}